\documentclass[doublecol]{epl2} 

\usepackage{amsmath,amssymb,amsfonts}
\usepackage{lineno}
\usepackage{wasysym}

\newcommand{\ie}{\emph{i.e., }}
\newcommand{\reff}[1]{(\ref{#1})}
\newcommand{\eref}[1]{Eq.\reff{#1}}
\newcommand{\erefs}[1]{Eqs.\reff{#1}}
\newcommand{\figref}[1]{Fig.\ref{#1}}
\newcommand{\citer}[1]{Ref.\!\!\cite{#1}}
\newcommand{\citers}[1]{Refs.\!\!\cite{#1}}
\newcommand{\p}{\partial}

\newcommand{\np}{n_p}
\newcommand{\nb}{n_B}

\newcommand{\sumiu}{\sum_{r=1}^{N}}

\newcommand{\kj}{k_j}

\newcommand{\kz}{k_0}

\newcommand{\lj}{\ell_j}

\newcommand{\lz}{\ell_0}

\newcommand{\omp}{\omega_p}

\newcommand{\vb}{v_B}

\newcommand{\phij}{\varphi_{k_j}}

\newcommand{\phija}{\phi_{\ell_j}}

\newcommand{\phiza}{\phi_{\ell_0}}

\newcommand{\eps}{\epsilon_p}

\newcommand{\xibar}{\bar{\xi}}

\newcommand{\betaj}{\beta_{j}}
\newcommand{\etab}{\bar{\eta}}

\usepackage{color}

\title{On the viability of the single wave model for the beam plasma instability}
\shorttitle{On the applicability of the single wave model}

\author{N. Carlevaro\inst{1,2} \and G. Montani\inst{1,3} \and D. Terzani\inst{4}}
\shortauthor{N. Carlevaro, G. Montani, D. Terzani}

\institute{        
\inst{1}ENEA, FSN-FUSPHY-TSM, R.C. Frascati, Via E. Fermi, 45 (00044) Frascati (RM), Italy.\\
\inst{2}L.T. Calcoli, Via Bergamo, 60 (23807) Merate (LC), Italy.\\
\inst{3}Department of Physics, ``Sapienza'' University of Rome, P.le Aldo Moro, 5 (00185) Roma, Italy.\\
\inst{4}Department of Physics, University of Naples ``Federico II'', Via Cinthia, I-80126 Napoli, Italy.
}

\abstract{We analyze the interaction of a cold fast electron beam with a thermalized plasma, in the presence of many Langmuir modes. The work aims at characterizing the deviation of the system behavior from the single mode approximation, both with respect to a consistent spectral analysis of the most unstable mode harmonics and with respect to the presence of a dense spectrum, containing linearly unstable and stable modes. We demonstrate how, on the one hand, the total energy fraction absorbed by the harmonics is negligible at all (by evaluating its total amount) and, on the other hand, the additional Langmuir modes can be excited via an avalanche mechanism, responsible for a transport in the particle velocity space. In particular, we show that the spectral broadening outlines a universal shape and the distribution function, associated to the avalanche mechanism, has an asymptotic plateau, differently from the coherent structures characterizing the single wave model.}

\begin{document}
\maketitle

\section{Introduction}
The understanding of the dielectric structure of a homogeneous plasma and of the existence of a stochastic electrostatic background of Langmuir waves \cite{Landau46} raised interest in the possibility to excite specific modes via the inverse non-linear Landau damping \cite{oneil65,mazitov65}. Indeed, the study of how a fast electron beam interacts with a thermalized plasma has received increasing contributions leading to a solid theoretical paradigm \cite{ZCrmp}.

A basic achievement must be regarded in the derivation of the linear dispersion relation for the beam-plasma system in \citer{OM68}, where a topological change of the Landau solution, due to the presence of a fast population, has been shown to imply the existence of unstable modes. The non-linear saturation of such beam-plasma instability and its backreaction on the fast particles has been successfully investigated in \citer{OWM71} (see also \citers{OL70,SS71,Ma72} and \citers{MK78,TMM94,AEE98,AJFR98,EEbook} for a rigorous generalized Hamiltonian reformulation of the problem). This latter work describes the growth and saturation of the Langmuir waves, as effect of the energy and momentum they receive from the fast particles when the resonance condition is nearly fulfilled, \ie when the beam velocity is almost equal to the phase velocity of the wave. The amplification of such a resonant mode is associated with a trapping process of fast particles. This paradigm, originally developed to interpret real experiments of beam-plasma interactions, is today of relevant interest in view of the possibility of implementing the ideas underlying the works by O'Neil and collaborators into the so-called bump-on-tail paradigm, \ie a warm tail in the electron plasma population (this topic is recently reviewed in \citer{BS11}). In fact, the bump-on-tail scenario is relevant for the interpretation of fusion oriented experiments, by a one-to-one correspondence between the Langmuir and shear Alfv\'en waves and between the transport in the velocity and radial spaces, respectively \cite{BB90a,CZ07,BS11,CZ13}. A discussion concerning the beam-plasma instability in the presence of a warm beam has been addressed in \citer{L72}, showing how the resonant mode is amplified as far as the relative velocity fluctuation is of the order of the ratio between the beam and the plasma density to the $1/3$. Both \citers{OWM71} and \cite{L72} actually deal with a single resonant wave model, properly justified as far as the beam is sufficiently tenuous.

In the present paper, we relax such a restriction and we consider a large number of Langmuir modes interacting with a single cold beam, with the aim to quantitatively test the validity of the single wave approximation. As a first step, we add to the resonant (most unstable) mode a wide number of harmonics, which are naturally generated when the beam heats and unavoidably acquires a certain degree of inhomogeneity. We study this problem to determine the consistency level of the harmonic inclusion, which requires to add hundreds of modes. The main merit of this analysis is to demonstrate how the total energy contribution of the harmonics is less than $10^{-5}$ times the resonant mode one. This confirms the ideas discussed in \citer{OWM71}, but with a quantitative self-consistent analysis of the Fourier coefficients during the beam charge distribution evolution.

Then, the core of the paper is devoted to a quantitative study of the regimes for which the single wave model approximation is broken. In particular, by enhancing the periodicity length of the system and varying the beam density, we determine the influence of a set of Langmuir modes (having a sufficiently large spectral density) on the profile of the resonance. We show how, when the beam density (or equivalently the mode spectral density) increases enough still remaining tenuous, it is possible to observe an avalanche process in which all the available linearly stable modes with larger wave numbers are also excited (see also \citer{FLA06}): this is due to the conservation of total momentum, since particles donate, on average, energy to the modes. In fact, it has been shown both theoretically (see, for example, \citer{TMM94}) and experimentally \cite{DEM05} that wave-particle interaction takes place locally in velocity space, a feature which is a source for the avalanche process mentioned above. As a result, in such a limit, the spectrum acquires a universal form (\ie we observe a universal line broadening) and the particle distribution function resembles a plateau profile, very different from the coherent structures observed in the phase space for the single wave model. Our study has the merit to clarify how the presence of a discrete Langmuir spectrum around a beam-plasma resonance can be very important for the transport properties in the velocity space and how, in this respect, the spectral density is crucial when compared with the beam intensity, indeed determining the universal character of the spectral intensity.

\section{Basic assumptions and linear analysis}

Let us now briefly recall the basic results of \citer{OM68}, where it is discussed the linear interaction of a single electron-beam with a cold background plasma treated as a one-dimensional (1D) dielectric medium supporting longitudinal electrostatic waves of frequency $\omega$. The bulk plasma is assumed homogeneous having a constant particle density $\np$, and the supra-thermal tenuous beam (having number density $n_B\ll\np$) has initial velocity $v_B$ much greater than the thermal electron velocity. The plasma frequency is defined as $\omp=\sqrt{4\pi\np e^2/m_e}$ ($m_e$ and $e$ being the unitary electron charge and mass, respectively) and, in the limit of a cold plasma, the dielectric function $\eps$ takes the simple form $\eps=1-\omp ^2/\omega^2$. Considering an electric field perturbations of the form $e^{ikx-i\omega t}$, from the study of the dispersion relation \cite{LP81,OM68} the most unstable mode is a Langmuir mode with $\omega\simeq\omp$, characterized by a wave vector $\kz$ satisfying the resonance selection rule $\kz=\omp/v_B$ (which guarantees that beam particles move at the wave phase velocity $v_{ph}=\omp/\kz=\vb$). Introducing the fundamental parameter $\etab=(\nb/2\np)^{1/3}$, the profile of the most unstable mode growth rate $\gamma(k)$ has a peaked structure with a maximum $\gamma_L\simeq 0.7\,\etab\omp$ in correspondence to $\kz$ and an half-width roughly estimated as $|\Delta k|/\kz\simeq 1.7\,\etab$ \cite{OM68,L72}. Actually, all wave numbers smaller that $\kz$ are (in principle) linearly unstable but, as we will see in the next Sections, they are substantially not influent in the non-linear dynamics due to total momentum conservation, and can be safely neglected in this estimate. Furthermore, we note that the dispersion relation is expanded around $\kz$ and so that it holds locally only. The Langmuir spectrum can thus be characterized as follows: the most unstable mode $\kz=\omp/\vb$; linear unstable modes having $\kz-|\Delta k|\apprle k \apprle \kz+|\Delta k|$; and the linear stable part with $k$ outside the instability region width.

\section{Non-linear beam-plasma interaction} 
Aiming at describing the trapping of the fast particles within the global electric profile and the generated evolutive spectrum, we follow the ``single wave model'' analysis in \citer{OWM71}, where only the most unstable mode has been addressed (see \citers{MK78,TMM94,AEE98,AJFR98,EEbook} for the Hamiltonian approach). Actually, we generalize the beam plasma dynamics to the presence of the linear unstable and stable part of the spectrum (details on the derivation of the system equations can be found in \citers{CFMZJPP} and \cite{CENTR}).

The 1D motion along the $x$ direction of the $N$ beam particles (located in $x_{r}$) is periodic of period $L$ and governed by the Newton law. The Poisson equation instead provides the self-consistent evolution of the mode, and the Langmuir wave scalar potential $\varphi(x,t)$ is expressed as a function of its Fourier components $\varphi_{k_{j}}(t)$ (here, $j=1,\,...,\,m$). Each mode has $\omega_j\simeq\omp$ and, in the assumption of a cold plasma, the dielectric function is nearly vanishing so that it can be formally expanded as $\eps\simeq (2i/\omp)\p_t$: this yields the evolution of the electric potentials. The dynamics is analyzed in the reference frame comoving with the initial beam speed, by introducing the effective non-linear shift $\xi$ defined as $\xi_r(t)=x_{r}-\vb t$. Moreover, the following set of normalized scaled quantities are introduced
\begin{equation}\label{adimvar}
\begin{split}
&\lj=\kj L/2\pi\;,
\;\;\;\;\;\;\;\;\;\;
\xibar_r=2\pi\xi_r/L\;,
\;\;\;\;\;\;\;\;\;\;
\tau=t\omp\etab\;,\\
&\phija=\phij\,e\kj^2/(m_e\etab^{2}\omp^2)\;,
\;\;\;
\betaj=(\kj \vb-\omp)/\omp\etab\;,
\end{split}
\end{equation}
where $\ell_j$ are integer numbers. The dynamical system governing the interaction of $m$ Langmuir modes and the injected beam reads now
\begin{subequations}\label{paijsdfjsoaisj}
\begin{align}
&\xibar_r''=i\;\sum_{j=0}^{m-1}\lj^{-1}\;\phija\;e^{i\lj\xibar_r+i\beta_j\tau}+c.c.\;,\label{poieqadi}\\
&\phija'=\frac{i}{N}\sumiu e^{-i\lj\xibar_r-i\betaj\tau}\;,\label{poieqadi2}
\end{align}
\end{subequations}
where the prime denotes the $\tau$ derivative. We mention how the present dynamics is formally equivalent to many Hamiltonian models studied over the last forty years (\cite{EEbook,AEE98,AEFR06,FE00,Farina04,Farina94,Duccio05,CFGGMP14}).

\section{Simulation results}
Here, we numerically explore the regime of validity for the single wave model. In particular, we study the effects of increasing the periodicity length of the system $L$ (or equivalently the value of $\etab$ and, thus, $|\Delta\ell|$) by considering $\lz\gg1$ and the possibility of avalanche excitation of the linear unstable and stable modes. The system \reff{paijsdfjsoaisj} is simulated using a $4$th-order Runge-Kutta algorithm (whose reliability is comparable to standard symplectic approaches, see for example \citers{HL03} and \cite{Ca93}) and $N=10^5$ total particles; for the considered time scales and for an integration step $\Delta\tau=0.01$, both the total energy and momentum (for the explicit expressions, see \citer{CFMZJPP}) are conserved with relative fluctuations of about $1.4\times10^{-5}$.

\begin{figure}[!ht]
\centering
\includegraphics[width=.6\columnwidth,clip]{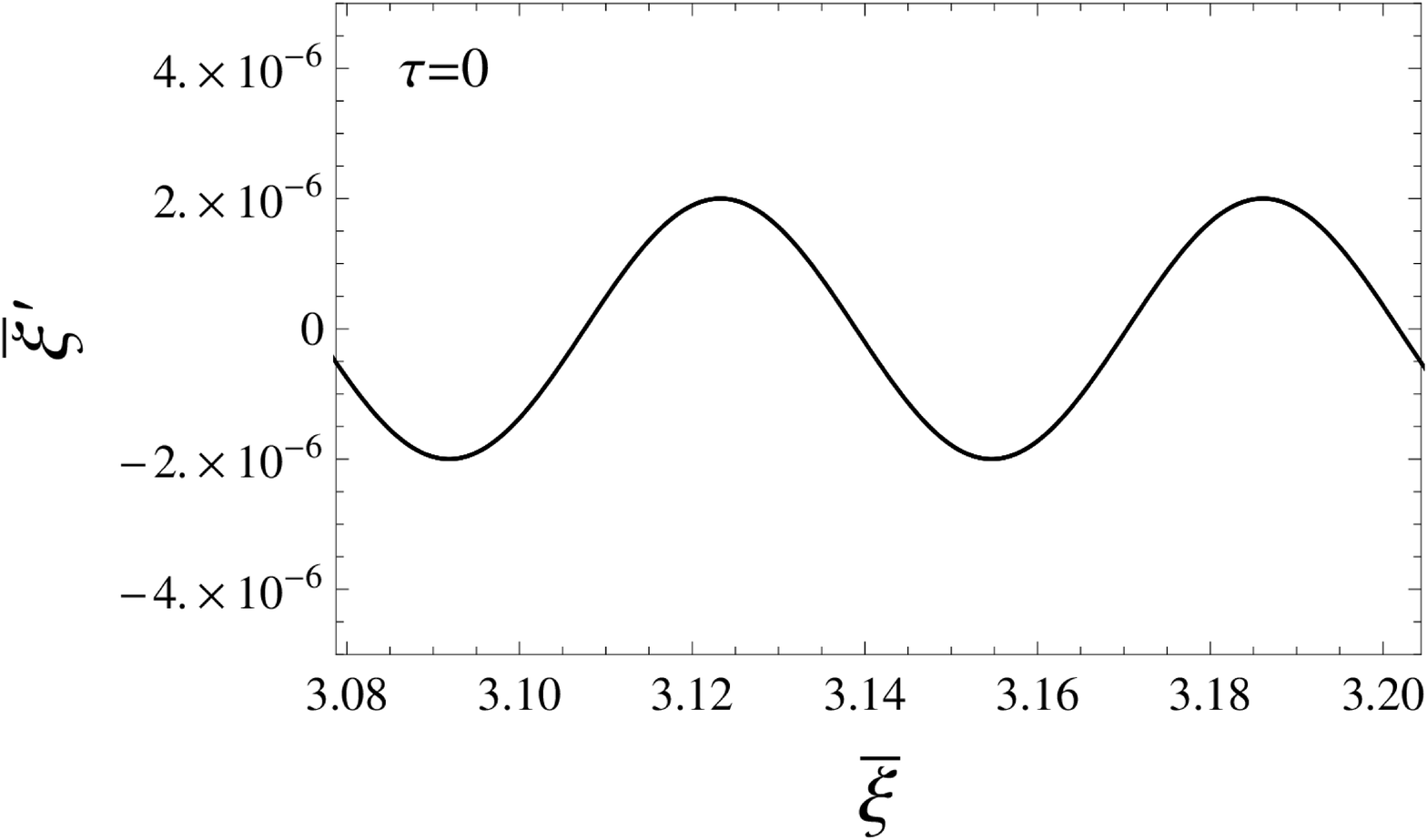}
\vspace{-1mm}
\caption{Zoom of the periodic initial conditions for $\bar{\xi}_{r}$ and $\bar{\xi}_{r}'$ in the restricted region near $\bar{\xi}=\pi$.}
\label{incond}
\end{figure}

The non-linear shifts $\bar{\xi}_{r}$ are normalized in \erefs{paijsdfjsoaisj} in the domain $[0,\,2\pi]$ and this feature is implemented in the code outputs.
The initial conditions are ideally assigned such that $\bar{\xi}_{r}(0)$ take random values while $\bar{\xi}_{r}'(0)$ are simply vanishing. But, accordingly to \citer{OWM71}, we set initial data (depicted in \figref{incond} around $\bar{\xi}=\pi$) slightly modified with respect to such a configuration in order to match the right value of the frequency predicted by the linear theory.
The initial values of the dimensionless complex modes are set with random phases and with mode amplitude of $|\phi_{\ell_{0}}(0)|=10^{-2}$ for the most unstable mode, and of $|\phi_{\ell_{j}}(0)|=10^{-4}$ for the other ones with $j\neq0$. The simulations are run implementing the transformation $\bar{\phi}_{\ell_{j}}=\phi_{\ell_{j}}e^{i\beta_j\tau}$ and the evolutive equations for the real and imaginary part of $\bar{\phi}_{\ell_{j}}$, respectively. It is worth mention that the frequency mismatch $\beta_j$ assumes the dimensionless expression $\beta_j=(\ell_j/\ell_0-1)/\etab$, where we have used the resonance condition $v_B=\omp L/(2\pi\ell_0)$.

\subsection{Single wave model}
We underline that the single wave model equations \cite{OWM71} ($m=1$ and $\lz=1$) are not explicitly dependent on $\etab$ (since $\beta_0=0$). The dynamics, taking a universal form, consists of two different stages: in the early evolution (until $\tau \sim 6$) an exponential growth of the mode takes place and then, after the non-linear amplitude saturation, the particles get trapped and begin to slosh in the potential well, making the mode intensity to oscillate. The instant $\tau \sim 6$ is called spatial bunching because of the behavior of the particle spatial distribution. In particular, the initially uniform beam gets trapped in the instantaneous potential well and the spatial density tends to a peaked profile. On the other hand, the velocity distribution spreads. From energy conservation at mode saturation $|\phi^{sat}|\simeq1.1$, the half non-linear velocity spread can be evaluated as
\begin{align}\label{deltav}
\Delta v_{nl}=|v_{nl}-\vb|\simeq\etab\omp\sqrt{4|\phi^{sat}|}/\kz\simeq 2.1\etab\omp/\kz\;.
\end{align}
Using now the expression of the linear growth rate, we can write the relation $\Delta v_{nl}\simeq1.4\,\omega_B/\kz\simeq3\,\gamma_L/\kz$ (where $\omega_B\simeq\etab\omp\sqrt{2|\phi^{sat}|}\simeq2.2\gamma_L$ is the bouncing frequency of trapped particles). This estimate, using scaled variables, simply provides $|\bar{\xi}'_{nl}|\simeq\sqrt{4|\phi^{sat}|}/\lz\simeq2.1$, as properly obtained from the numerical simulations.

\subsection{Harmonics of the most unstable mode}
Let us now deal with the problem of the intrinsic validity of the single wave model in the presence of harmonics of $\lz$. This issue was already discussed in \citer{OWM71} and here we deepen the analysis by studying the possible effects on the long time dynamics of the whole harmonic Fourier spectrum. In fact, in \citer{TMM94} it was pointed out that the first harmonics are negligible only during the very early stages of motion because of the initial spatial bunching. At saturation, the system exhibits a peaked profile in the $\bar{\xi}$ space and the Poisson equation shows that every harmonic of the fundamental mode is proportional to the Fourier coefficient of the density $\hat{\rho}_{k_j}$: thus, any of them could be excited. In the following, we study the presence of the harmonics of the most unstable mode: this corresponds to implement the dynamics of wave numbers $k_h=\omega_h/(v_{ph})_h=h\omp/\vb=h\kz$, where we have introduced the integer harmonic index $h=2,\,...,\,p$. The dielectric function is not vanishing since $\omega_h=h\omp$, and the evolution of an harmonic field is thus governed by an algebraic equation coming directly from the Poisson law, \ie
\begin{align}
\phi_{\ell_h}=-\frac{2\etab}{\eps(h\omp)N}\sum_{r=1}^N e^{-i\left(\ell_h\xibar_r+\beta_h\tau\right)}\;,
\end{align}
instead of a differential one (its modulus is reduced by $\etab$).

In order to demonstrate the long time validity of the single wave model (for the sake of simplicity, we set $\ell_0=1$ and $\etab=0.01$), we have firstly run a simulation with only the most unstable mode in order to extract all the coefficients $\hat{\rho}_{k_h}$ through a Fourier analysis. Then, we have simulated the system in the presence of the outlined relevant harmonics (about 100). In this latter case, also the harmonic modes are initialized as $|\phi_{\ell_h}(0)|=10^{-2}$ (as in \citer{OWM71}) and the simulation was performed until $\tau=200$.
\begin{figure}[!ht]
\begin{flushright}
\includegraphics[width=.49\columnwidth,height=2.96cm,clip]{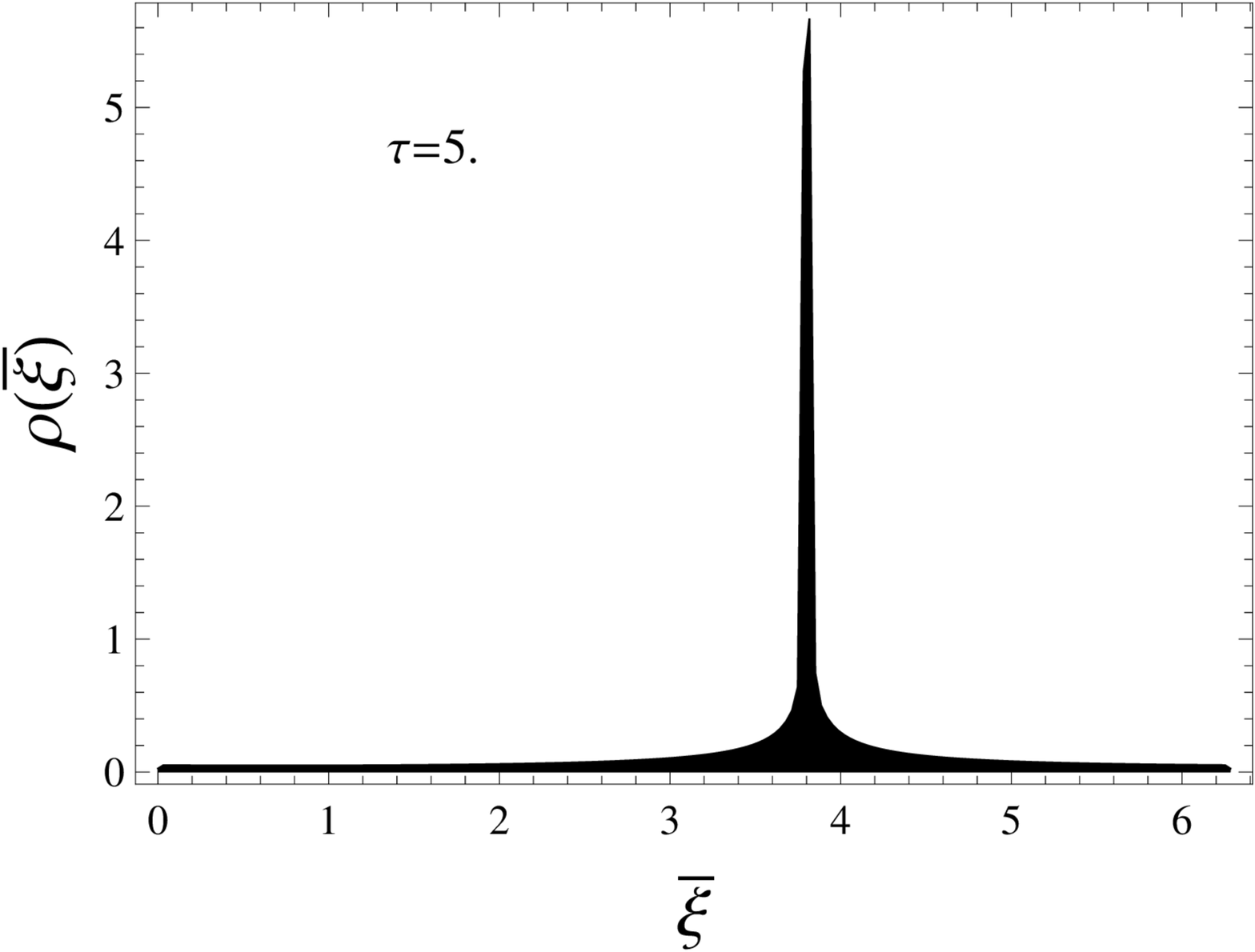}
\includegraphics[width=.49\columnwidth,height=3cm,clip]{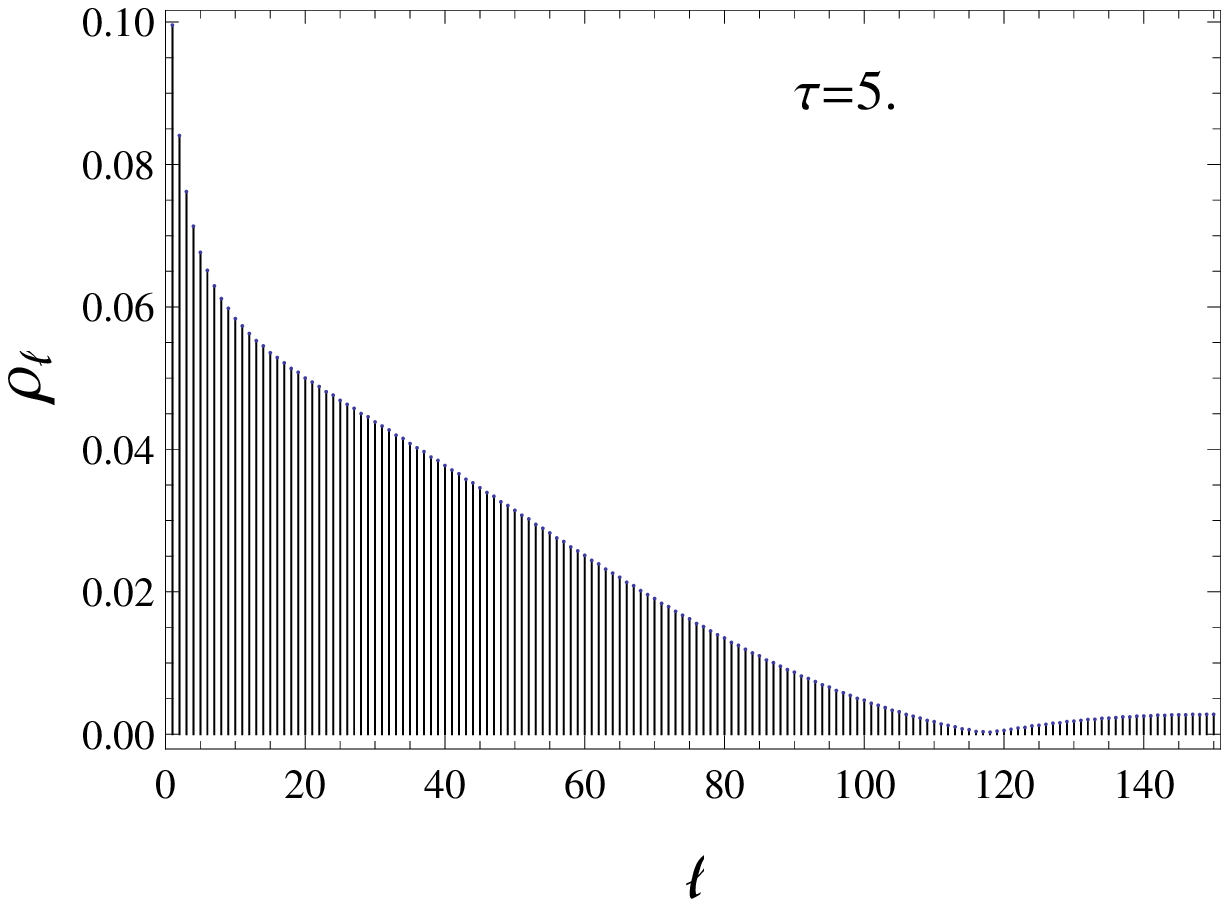}
\end{flushright}\vspace{-8mm}
\caption{Single wave model harmonic generation. Te left-hand panel shows the density $\rho(\xibar)$ near the spatial bunching; in the right-hand panel, the corresponding Fourier coefficients are computed and plotted: a threshold at the hundredth harmonic is evident.}
\label{swmexpansion}
\end{figure}
\begin{figure}[!ht]
\begin{flushright}
\includegraphics[width=.49\columnwidth,height=2.96cm,clip]{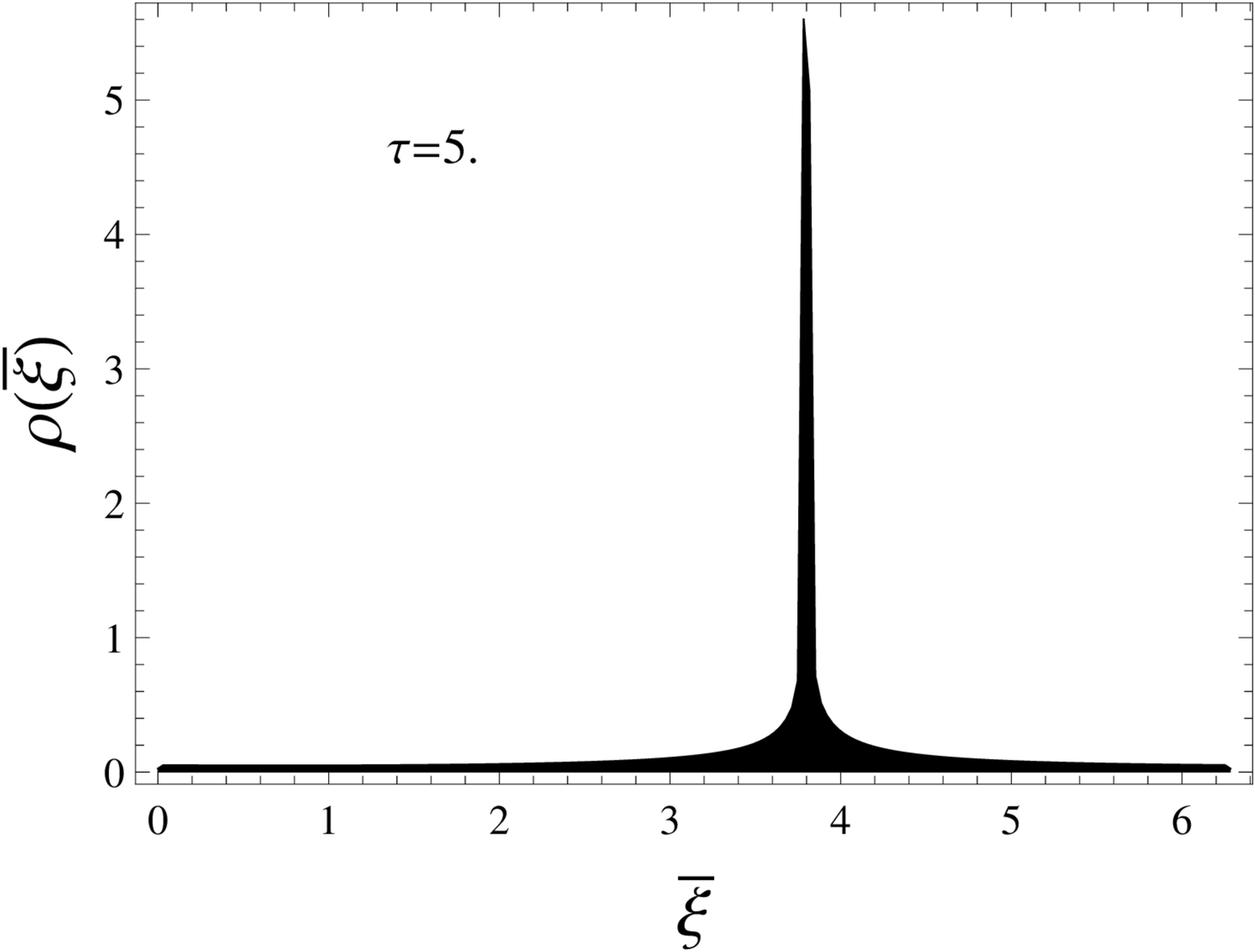}
\includegraphics[width=.49\columnwidth,height=3cm,clip]{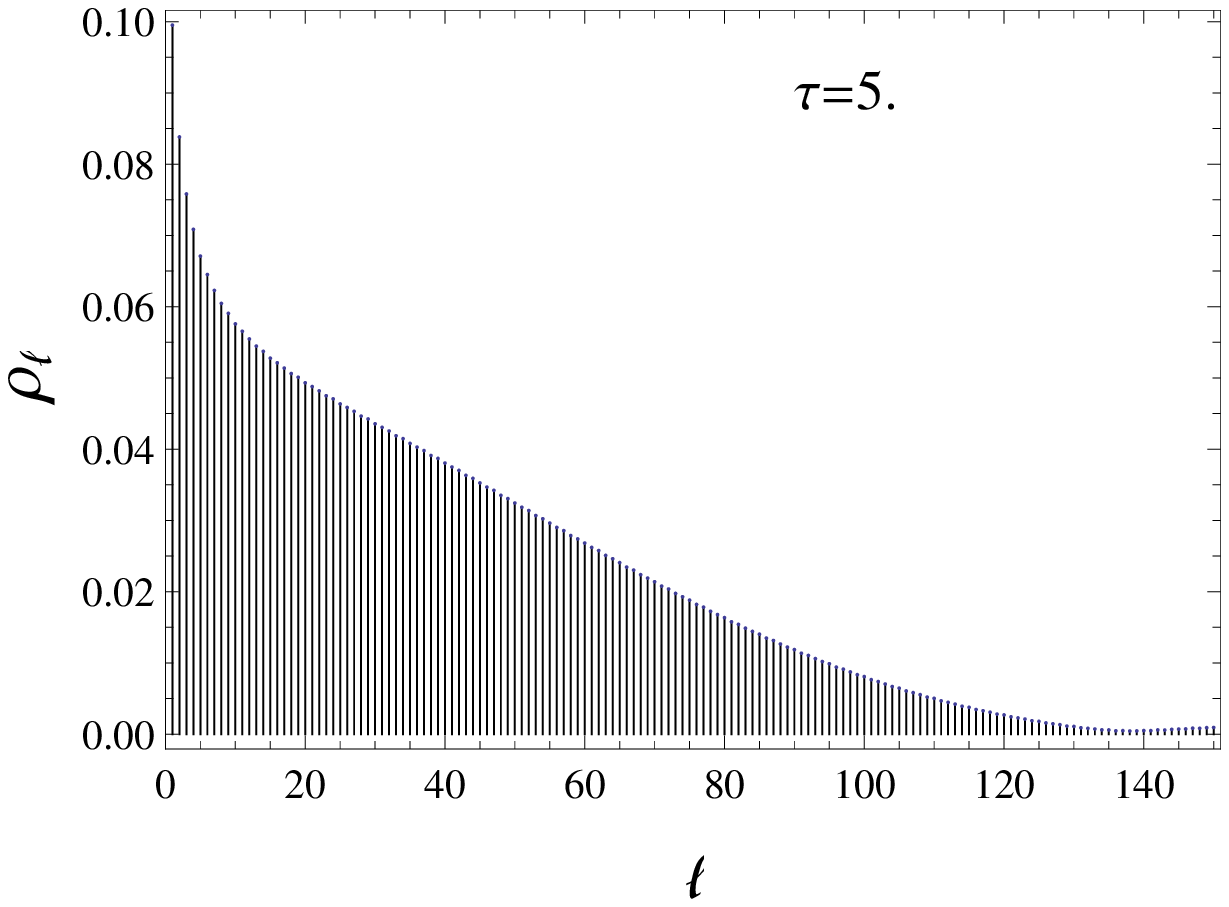}
\end{flushright}\vspace{-8mm}
\caption{Simulation in the presence of harmonics. The left-hand panel shows the density $\rho(\xibar)$ at spatial bunching in the presence of the first 100 harmonics: the dynamic results are unaltered and the production of harmonics is unchanged (right-hand panel).}
\label{mwexpansion}
\end{figure}

In \figref{swmexpansion}, the single wave model harmonic generation is outlined and the density $\rho(\xibar)$ (left-hand panel) and its Fourier coefficients (right-hand panel) are shown at $\tau=5$ (almost at spatial bunching). From this analysis, it is clear how only the first hundred harmonics can be assumed as non negligible, so we have considered those hundred additional modes in the original system. The results of the extended simulation are shown in \figref{mwexpansion} and it emerges how the hypothesis for which the harmonics do not influence the system is consistently verified since no quantitative changes in the dynamics occur.
\begin{figure}[!ht]
\centering
\includegraphics[width=.5\columnwidth,clip]{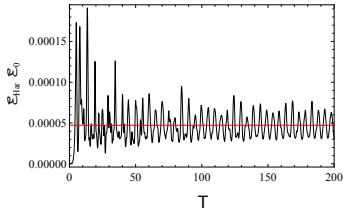}
\vspace{-4mm}
\caption{Evolution of the ratio $\mathcal{E}_{Har}/\mathcal{E}_0$; the red line represents the mean value $\simeq5\cdot 10^{-5}$. (Colors online)}
\label{energyratio}
\end{figure}
To corroborate this observation, we now analyze the harmonic evolution for long times by studying the energy ratio that they subtract from the leading mode $\phiza$. We monitor the ratio $\mathcal{E}_{Har}/\mathcal{E}_0=|\ell_0\phiza|^{-2}\sum_h |\ell_h\phi_{\ell_h}|^2$, that provides the energy percentage lost by $\phiza$. We plot the temporal evolution of this quantity in \figref{energyratio}, where we also indicate (in red) the time average $\simeq5\times10^{-5}$ (we underline how $\mathcal{E}_{Har}/\mathcal{E}_0$ scales like $\etab^2$, so that for the upper-limit value $\etab\simeq0.2$ it could take the value $4\times10^{-3}$). In conclusion, this whole analysis strengthens and quantitatively characterizes the assumption that the presence of harmonics can be safely neglected in the single wave model. We emphasize that the short time analysis is needed to ensure the self-consistency of the considered scheme (\ie new harmonics are not significantly generated), while the long time study sheds light on the average negligible contribution to the particle motion due to harmonics.

\subsection{Analysis of the multi-mode system}
Let us now analyze the effects of the linear unstable and stable part of the spectrum. In particular, we simulate the non-linear dynamics involving a large set ($m=60$) of Langmuir modes.

For the single wave model, comparing the estimate of the non-linear velocity spread \eref{deltav} and the expression of the linear instability width, one can easily recognize that $\Delta v_{nl}/\vb\simeq1.2\,\Delta k/\kz$. This indicates how the linear unstable modes should also feel the drive of particles that have been spread in velocity due to non-linear interaction.
\begin{figure}[!ht]
\centering
\includegraphics[width=.7\columnwidth,clip]{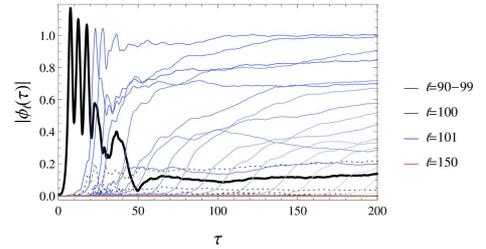}
\vspace{-5mm}
\caption{Evolution of mode amplitudes with $90\leqslant\ell\leqslant150$ (here we set $\etab=0.01$). A sort of avalanche effect is evident, for which the most unstable mode $\ell=100$ (thick black line) passes its energy to the following ones (dashed lines). The colors follow the temperature map sequence, from blue ($\ell=101$) to red ($\ell=150$). The mode numbers smaller than $\lz$ (dotted black lines) are not destabilized at all. (Colors online)}
\label{mwphi}
\end{figure}
A multi mode model is therefore mandatory, at fixed $\etab$, for $\ell_0\gg1$, thus enlarging the unstable spectrum $\Delta\ell$. From $\Delta\ell/\ell_0\simeq1.7\etab$, it can be argued that the single wave approximation is valid only for $\etab\lz\ll1$. Moreover, we will point out that the non-linear particle velocity spread is increased by the unstable modes through non-diffusive particle transport and can also excite the linear stable part of the spectrum, triggering a sort of avalanche effect. In fact, the excited modes exchange energy with the beam, so its spread in velocity increases destabilizing the nearest mode and so on. In this case, a highly dense spectrum can be excited by a single fast beam, underling the importance the linear stable modes have on the transport features.

Let us now discuss the simulation results of a system with $\lz=100$ and $\etab=0.01$\footnote{It is worth noting that the treated processes present a time scale that is invariant for changes of $\ell_0$. In fact, saturation occurs always after $\tau \sim 5$, even though $\ell_0$ is changing. Moreover, in the motion equation, any term with the product $\ell_j \tau$ is present and running simulations decreasing the time step by two order of magnitude gives a perfect match in the results.}. The number of modes effectively participating in the dynamics has been chosen, after some tests, to be $60$. Thus, we run simulations with $90\leqslant\ell\leqslant150$, where the asymmetry with respect to $\lz$ is due to the fact that the particles lose energy (decreasing their velocity) and can thus excite large wave numbers. In \figref{mwphi}, the behavior of the whole Langmuir set is shown as a function of time. It is evident that initially the system behaves as a single wave model, but, after the excitation of the adjacent linear unstable modes, the avalanche effect begins and stable modes are also non-linearly destabilized. Thus, through transport of particles to lower velocity, modes with large wave number acquire a non-linear drive. This analysis underlines the relevance of the linear unstable modes for the case $\lz\gg1$ and enlightens the failure of the single wave model assumption.
\begin{figure}[!ht]
\centering
\includegraphics[width=.49\columnwidth,clip]{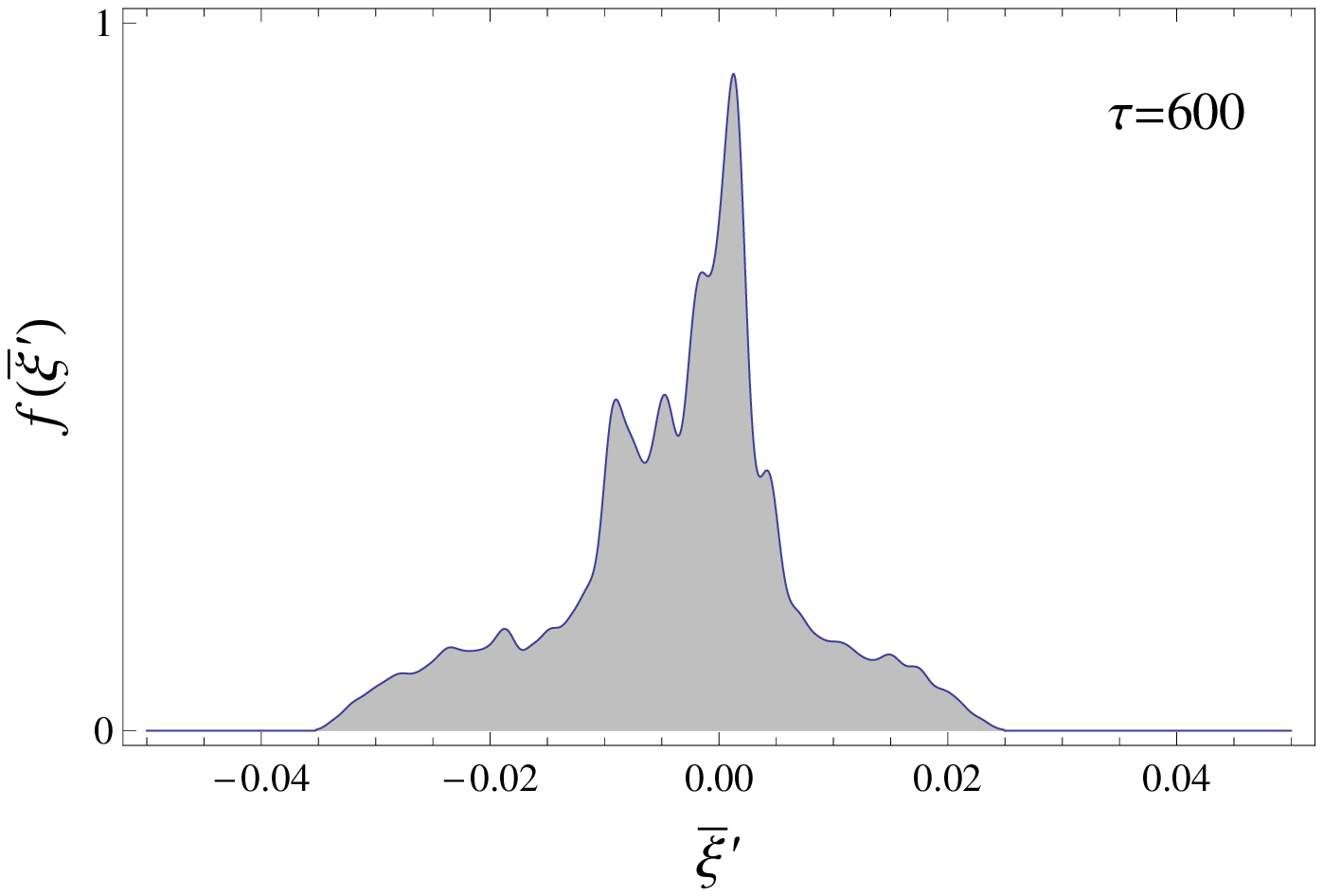}
\includegraphics[width=.49\columnwidth,clip]{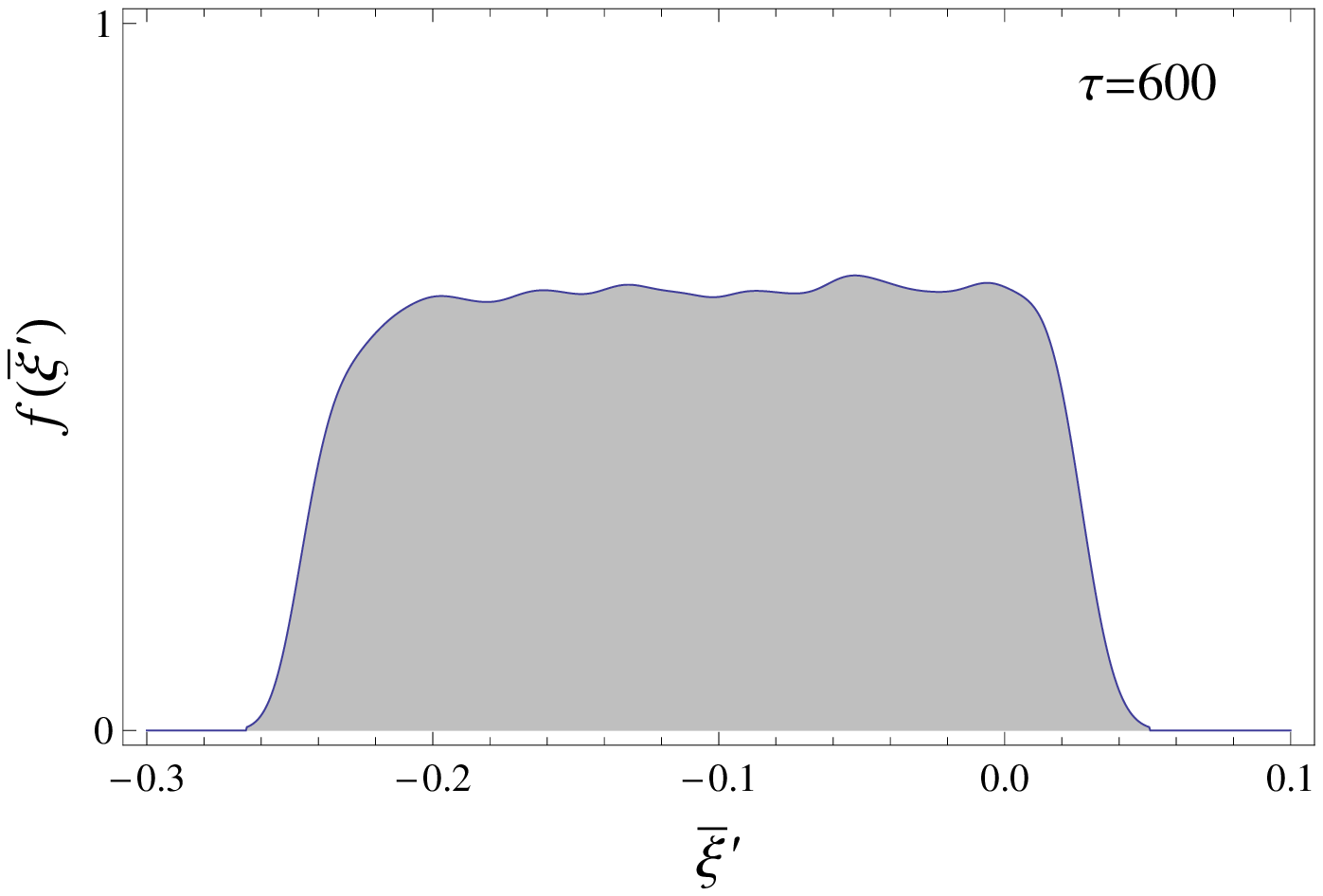}
\vspace{-9mm}
\caption{Comparison between the velocity distribution in the single wave model (left-hand panel) and in the multi-mode system (right-hand panel). In the first case, the presence of a macroparticle is pointed out by the peaked profile; in the second case, the multi-mode interaction flattens the distribution leading to a plateau.}
\label{pdfconfront}
\end{figure}

The avalanche excitation phenomenon is described by the destruction of the coherent features of the clump profile. In fact, an almost uniform particle distribution in $\mu-$space emerges instead of the rotating structure pointed out in the single wave assumption. We may better characterize this effect analyzing the velocity distribution $f(\bar{\xi}')$ in comparison with the single wave results (suitably normalized). In \figref{pdfconfront}, a deep difference between the two cases emerges. In fact, in the left-hand panel, a peaked profile is outlined corresponding to the presence of the rotating clump, that oscillates indefinitely in time and describes the particle trapping phenomenon. On the other hand, in the right-hand panel, the formation of a \emph{plateau} in the velocity distribution occurs: this is due to the fact that the interaction with the linear stable spectrum does not allow the confinement of the particles in coherent structures. In fact, particle transport and diffusion take place leading to a loss of mean velocity of the beam and to a consequent flattening of the distribution profile.

\subsection{Spectrum evolution}
\begin{figure}[!ht]
\centering
\includegraphics[width=.49\columnwidth,clip]{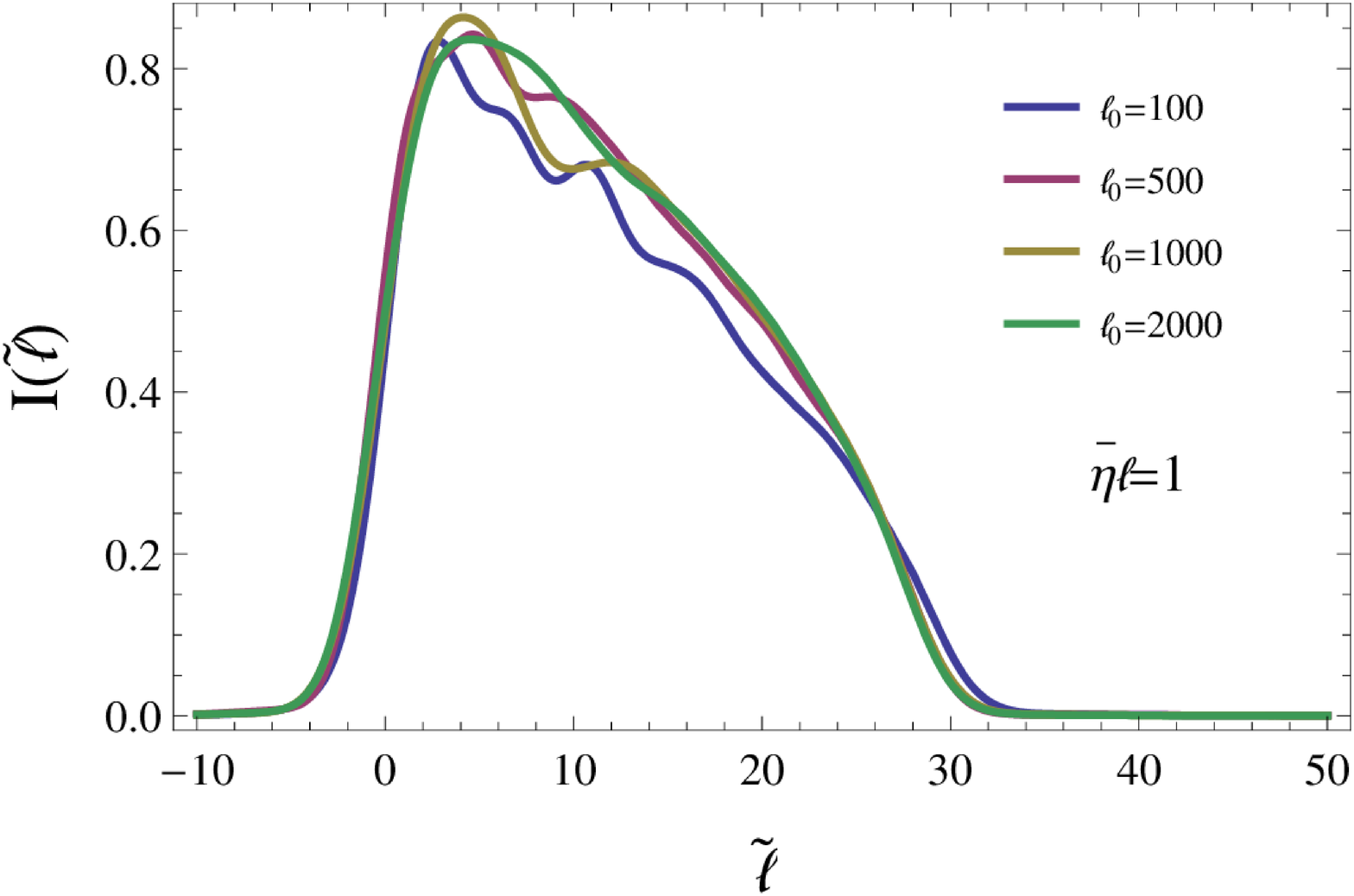}
\includegraphics[width=.49\columnwidth,clip]{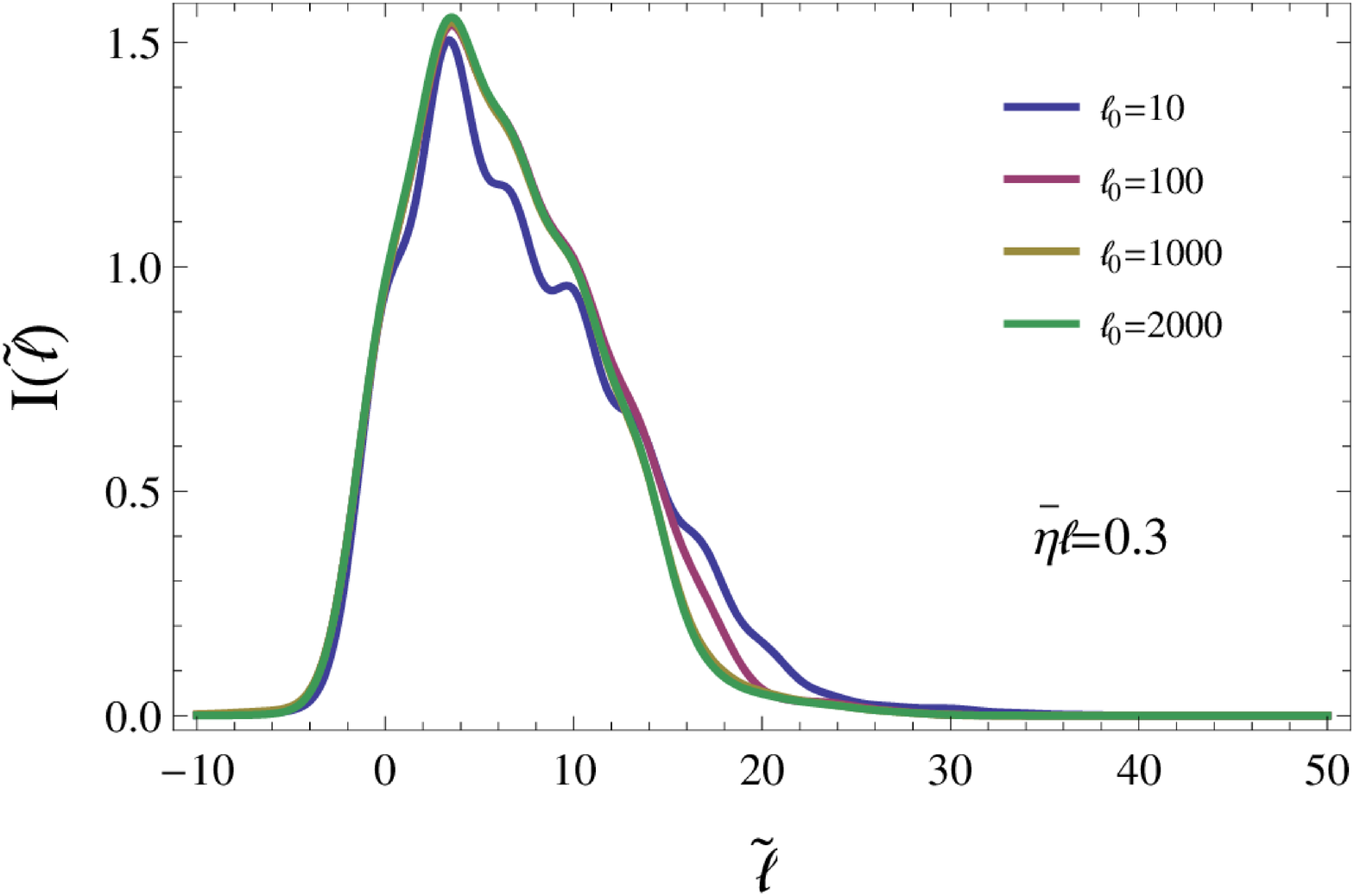}
\vspace{-5mm}
\caption{Superposition of the curves $I(\tilde{\ell})$ at late times, for constant $\etab\ell_0=1$ (left-hand panel) and $\etab\ell_0=0.3$ (right-hand panel). The values of $\ell_0$ are indicated in the plots. A universal spectral profile emerges, with a good degree of accuracy, for the classes $\etab\ell_0=const.$ (Colors online)}
\label{idik}
\end{figure}

Let us now define the quantity $I(\ell_j)=|\phi_{\ell_j}|^2$ and let us analyze the behavior of some fundamental parameters of its evolution, such as the peak position and the decay slope as function of the $j$ index. As a result, we point out how the spectral curve $I(\ell_j)$ takes an asymptotic universal profile as the parameter $\etab\lz$ remains constant. In this respect, we analyze two sets of four cases corresponding to $\etab\lz=1$ and $\etab\lz=0.3$, respectively. By scaling the wave number as $\tilde{\ell}=(\ell - \ell_0)/(\etab \lz)$, the curves corresponding to the four cases asymptotically behave as in \figref{idik}, left(right)-hand panel for $\etab\lz=1(0.3)$, where a curve was assumed to be asymptotic when the modes with very large $\ell$ no longer vary significantly (after some tests, the profiles were taken for temporal scales of order $\mathcal{O}(500)$).
The four curves consistently overlap as a demonstration of the universal behavior of the system in function of the parameter $\etab\lz$. The breaking of the single wave model approximation is, furthermore, evident in a system with $\etab\ell_0\sim1$, while it properly holds for $\etab\ell_0\ll1$.

These two results find a clear explanation as far as we discuss how the quantity $\etab\ell_0$ enters the non-linear physics of the model. As discussed above, we see that the resonance width reads $\Delta\ell\simeq1.7\etab\ell_0$, thus the universal character of the spectral profile is associated with a fixed number of linear unstable modes. Furthermore, since $\Delta\ell$ must be at least one, once fixed $\ell_0$ a value of $\etab$ (such that $\etab\ell_0\ll1$) always exists for which the non-linear velocity spread $\Delta v_{nl}/v_B\simeq2\etab$ is too small to excite any additional mode but the most unstable one.

The time evolution of $I(\tilde{\ell})$ is characterized by two stages, as can be argued from \figref{spettrtime}:
\begin{figure}[!ht]
\centering
\includegraphics[width=.6\columnwidth,clip]{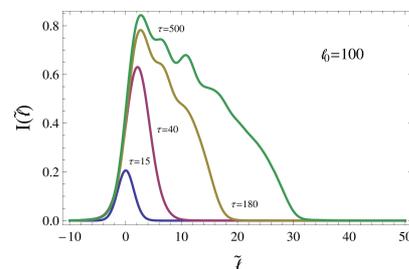}
\vspace{-5mm}
\caption{Temporal evolution of the spectral profile $I(\tilde{\ell})$ for the case $\etab\ell_0=1$, with $\ell_0=100$. Each curve corresponds to a different time as indicated in the plot. (Colors online)}
\label{spettrtime}
\end{figure}
first, most of the energy of the leading mode is transferred to the adjacent ones, and, after the generation of a peaked spectrum, a small shift of such a profile towards larger wave-numbers is present; then a phase of evolution starts in which the amplitude of the leading modes does not vary considerably, but the support of the function increases, denoting the excitation of increasing large $\ell$.

\section{Concluding remarks}
This paper was devoted to deepen the understanding of the beam-plasma interaction in the presence of many Langmuir waves. In particular, we included in the system dynamics both the harmonics of the most unstable mode (with frequency multiple of the plasma one) and an array of wave numbers, linearly unstable and stable for that parameter range. We demonstrated that, as far as the product $\etab\lz$ is much less than unity, the single wave model is consistent. However, when such a product increases enough, the modes having a wave number greater than the most unstable one are also excited due to the beam non-linear broadening in the velocity space and an avalanche phenomenon can take place, \ie the particle transport is significantly enhanced, leading to the formation of a plateau profile in the velocity distribution function as the asymptotic evolution is concerned. This result underlines the relevance of the spectrum morphology in determining the proper evolution of the beam-plasma instability and how the two crucial parameters are the beam intensity and the mode spectral density.

In addition to this basic issue of our study, we demonstrated, on the one hand, how the energy fraction absorbed by the harmonics of the most unstable mode is negligible at all; on the other hand, that the non-linear broadening of the spectrum possesses a universal morphology, depending on the value of the product $\etab\lz$ only. In the light of a weak turbulent plasma theory \cite{Du66}, for which arbitrarily high Fourier components enter the system dynamics, the present study suggests that the single wave model can not predict the transport feature  when isolated resonances are overlapped, as in physical systems.

We stress how in \citers{LP99} and \cite{BEEB11}, the validity of the so-called quasi-linear model was investigated, and the assumptions under which such a paradigm is predictive were clarified. In particular, in \citer{BEEB11}, it is shown the absence of the mode coupling in the saturation mechanism of the bump-on-tail instability. Also the present analysis outlines that, in the meso-scale temporal evolution of the system, avalanche processes can take place, and we clarify how, in such a regime, convection phenomena are comparably relevant as diffusion ones, see also \citer{CENTR}. By other words, the spectral intensity evolution is associated to coherent features which shift the peak of the spectrum preserving the resonance selection rule. However, this phenomenon coexists with a diffusion process, dominating the late evolution phases of the system and being responsible for the plateau profile in the particle distribution function.

\acknowledgments{\footnotesize We would like to thank Fulvio Zonca for his valuable suggestion on these topics, especially about the existence of a universal spectral morphology. The work of NC and GM has been carried out within the framework of the EUROfusion Consortium as project ER15-ENEA-03 (``NLED''). The views and opinions expressed herein do not necessarily reflect those of the European Commission.}

\end{document}